\documentclass[pra,aps,amsmath,amssymb,amsfonts,twocolumn,nofootinbib,showpacs,superscriptaddress, ]{revtex4}
\usepackage{bm,mathrsfs}

\usepackage{graphicx}
\usepackage{epsfig}
\usepackage{amsmath,bbm}
\usepackage{amsfonts,amssymb}
\usepackage{times}
\usepackage{verbatim}
\usepackage[sort&compress]{natbib}

\usepackage{amsmath}
\usepackage[colorlinks,breaklinks,linkcolor=blue,anchorcolor=blue,citecolor=blue,urlcolor=blue,
dvipdfm
]{hyperref}

\newcommand{\be}{\begin{equation}}
\newcommand{\ee}{\end{equation}}
\newcommand{\bea}{\begin{eqnarray}}
\newcommand{\eea}{\end{eqnarray}}

\newcommand{\ket}[1]{|#1\rangle}
\newcommand{\bra}[1]{\langle#1|}

\def\>{\rangle}
\def\<{\langle}

\def\qed{\leavevmode\unskip\penalty9999 \hbox{}\nobreak\hfill
     \quad\hbox{\leavevmode  \hbox to.77778em{%
               \hfil\vrule   \vbox to.675em%
               {\hrule width.6em\vfil\hrule}\vrule\hfil}}
     \par\vskip3pt}
     
\begin{document}
\newtheorem{theorem}{Theorem}
\newtheorem{lemma}[theorem]{Lemma}
\newtheorem{corollary}[theorem]{Corollary}
\newtheorem{proposition}[theorem]{Proposition}
\newtheorem{definition}[theorem]{Definition}
\newtheorem{example}[theorem]{Example}
\newtheorem{conjecture}[theorem]{Conjecture}

\title{{Dissipative preparation of steady Greenberger-Horne-Zeilinger states for Rydberg atoms with quantum Zeno dynamics}}

\author{X. Q. Shao\footnote{shaoxq644@nenu.edu.cn}}
\affiliation{Center for Quantum Sciences and School of Physics, Northeast Normal University, Changchun, 130024, People's Republic of China}
\affiliation{Center for Advanced Optoelectronic Functional Materials Research, and Key Laboratory for UV Light-Emitting Materials and Technology
of Ministry of Education, Northeast Normal University, Changchun 130024, China}
\affiliation{Department of Physics, Tsinghua University, Beijing, 100084, People's Republic of China}

\author{J. H. Wu}
\affiliation{Center for Quantum Sciences and School of Physics, Northeast Normal University, Changchun, 130024, People's Republic of China}
\affiliation{Center for Advanced Optoelectronic Functional Materials Research, and Key Laboratory for UV Light-Emitting Materials and Technology
of Ministry of Education, Northeast Normal University, Changchun 130024, China}

\author{X. X. Yi}
\affiliation{Center for Quantum Sciences and School of Physics, Northeast Normal University, Changchun, 130024, People's Republic of China}
\affiliation{Center for Advanced Optoelectronic Functional Materials Research, and Key Laboratory for UV Light-Emitting Materials and Technology
of Ministry of Education, Northeast Normal University, Changchun 130024, China}

\author{Gui-Lu Long\footnote{gllong@mail.tsinghua.edu.cn}}
\affiliation{Department of Physics, Tsinghua University, Beijing, 100084, People's Republic of China}
\date{\today}

\begin{abstract}
Inspired by a recent work [Reiter, Reeb, and S{\o}rensen, {\color{blue}Phys. Rev. Lett. {\bf117}, 040501 (2016)}], we present a simplified proposal for dissipatively preparing a Greenberger-Horne-Zeilinger (GHZ) state of three Rydberg atoms in a cavity. The $Z$ pumping is implemented under the action of the spontaneous emission of $\Lambda$-type atoms and the quantum Zeno dynamics induced by strong continuous coupling. In the meantime, a dissipative Rydberg pumping breaks up the stability of the state $|{\rm GHZ}_+\rangle$ in the process of $Z$ pumping, making $|{\rm GHZ}_-\rangle$ be the unique steady state of system. Compared with the former scheme, the number of driving fields acting on atoms is greatly reduced and only a single-mode cavity is required. The numerical simulation of the full master equation reveals that a high fidelity $\sim98\%$ can be obtained with the currently achievable parameters in the Rydberg-atom-cavity system.
\end{abstract}


\maketitle

\section{Introduction}

Neutral atoms have shown great potential as matter qubits that possess high-lying Rydberg states and state-dependent interaction. These properties make it possible to implement quantum information processing since the entangling operations can be readily realized by the Rydberg blockade or antiblockade interaction \cite{Jaksch2000,PhysRevLett.98.023002,Urban2009,gaetan2009observation,PhysRevLett.104.013001,Saffman2010}. There is currently great interest in generation of entangled states of Rydberg atoms using time-dependent unitary method. Theoretically, the multipartite entanglements were produced through stimulated Raman adiabatic passage \cite{PhysRevLett.100.170504} and asymmetric Rydberg blockade \cite{PhysRevLett.102.240502}, respectively, and a spatial cat state for a pair of atom clouds was created via the mechanism of Rydberg dressing \cite{PhysRevA.87.051602}. Experimentally, significant achievements have been obtained towards this field, e.g., using identical $^{87}$Rb atoms and $^{133}$Cs atoms, the deterministic Bell states with fidelities of $75\%$ and $82\%$ were demonstrated \cite{PhysRevLett.104.010502,PhysRevA.82.030306,Jau_Hankin_Keating_Deutsch_Biedermann_2015}. For non-identical particles, the entanglement between a $^{85}$Rb atom and a $^{87}$Rb atom via Rydberg blockade was reported as well \cite{arxiv}.

The reservoir-engineering approaches to entanglement generation have attracted much attention in recent years. In such methods, a detrimental source of noise can be converted into a resource, and the target state is the unique steady state of the open quantum system, which means there needs no state initialization.                                                                                                          Since the novel concept of ``quantum computation by dissipation"  was proposed by Verstraete {\it et al.} \cite{Verstraete2009}, the steady entangled states of two particles have been carried out numerously in various physical systems, including cavity QED systems \cite{Kastoryano2011,Lin2013}, ion trap systems \cite{Bentley2014}, optomechanical systems \cite{PhysRevLett.110.253601}, superconducting systems \cite{Leghtas2013,shankar2013}, and neutral atom systems \cite{Carr2013,PhysRevLett.111.033606,PhysRevA.89.052313}, etc. Nevertheless, it remains a challenge to prepare steady multipartite entanglement in a dissipative way.
    Recently, Morigi {\it et al.} put forward a protocol for dissipative quantum control of a spin chain, where an entangled antiferromagnetic state of many-body system was stabilized on the basis of spectral resolution, engineered dissipation, and feedback \cite{PhysRevLett.115.200502}. Subsequently, Reiter {\it et al.} present a scalable way for dissipative preparation of multipartite GHZ state without feedback \cite{PhysRevLett.117.040501}. In their scheme, a ``$Z$ pumping" and a ``$X$ pumping" constitute two crucial operations during the quantum-state preparation, and both of them require an independent harmonic oscillator mode and classical multitone driving fields operated on atoms. In particular, the preparation of steady GHZ state for $N$ particles has to involve $2(N-1)$ driving tones in the $Z$ pumping and  $2\lfloor(N+1)/2\rfloor$ driving tones in the $X$ pumping altogether. It may therefore consume many resources in terms of experimental realization.

\begin{figure}
\includegraphics[scale=0.20]{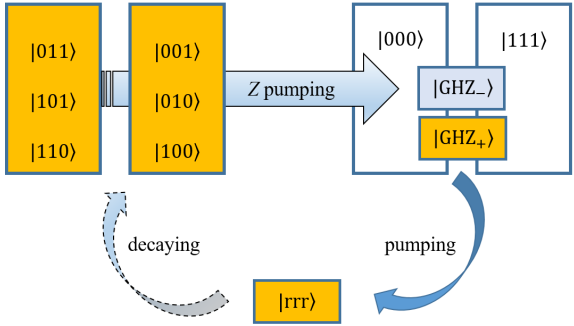}
\caption{\label{pa1}Protocol for preparation of the GHZ state: a simplified $Z$ pumping first transforms one and two atoms in state $|1\rangle$ into $|000\rangle=(|\rm{GHZ_+}\rangle\pm|\rm{GHZ_-}\rangle)/\sqrt{2}$, and in the eigenstates of the parity operator ${\cal P}=\Pi_{i=1}^3(|1\rangle_{ii}\langle0|+|0\rangle_{ii}\langle1|)$, the state $|\rm{GHZ_+}\rangle$ can again be rewritten in the form $|\rm{GHZ_+}\rangle=(|+++\rangle+|+--\rangle+|-+-\rangle+|--+\rangle)/2$. Then a resonant Rydberg pumping  couples to the transition from $|+++\rangle$ to $|rrr\rangle$, thereby the stability of state $|{\rm GHZ_+}\rangle$ under the $Z$ pumping is destroyed, leaving the unique steady state $|{\rm GHZ_-}\rangle$ unchanged. }
\end{figure}

In this work, we concentrate on the dissipative generation of tripartite GHZ state in a composite system based on Rydberg atoms and an optical cavity. The interaction between Rydberg atoms and cavity have been extensively studied before, e.g., a Rydberg-blocked atomic ensemble has a collective enhancement $\sqrt{N}$ coupling strength compared to the single atom as placed in an optical high-finesse cavity \cite{PhysRevA.82.053832}, and the Rydberg polaritons (a kind of quasiparticle with photons stored in the highly excited collective states) enable people to find new mechanisms of interaction in quantum optics \cite{PhysRevLett.110.090402,PhysRevLett.110.103001}.
The diagram of the protocol for preparing the GHZ state is illustrated in Fig.~\ref{pa1}. Similar to the process of Ref.~\cite{PhysRevLett.117.040501}, there are two operations to accomplish the goal, one is the $Z$ pumping that transforms one and two atoms in state $|1\rangle$ into $|000\rangle=(|\rm{GHZ_+}\rangle\pm|\rm{GHZ_-}\rangle)/\sqrt{2}$, and the other is the dissipative Rydberg pumping which induces a resonant transition between states $|+++\rangle$ and $|rrr\rangle$, and then rules out the steady population of state $|\rm{GHZ_+}\rangle$. In what follows, we will discuss in detail the feasibility of realization of the above operations in a Rydberg-atom-cavity system,  and it shows that our scheme can greatly reduce the complexities of experimental operations.

\section{physical system}
\begin{figure}
\includegraphics[scale=0.25]{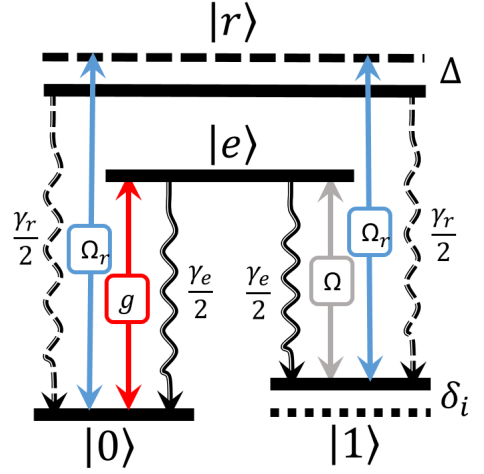}
\caption{\label{fig1} Schematic view of the four-level Rydberg atom. The quantum bit is encoded into the ground states $|0\rangle$ and $|1\rangle$. A quantized cavity mode is coupled to the transition between $|0\rangle$ and $|e\rangle$ with strength $g$, while a classical field of Rabi frequency $\Omega$ drives the atomic transition from $|1\rangle$ to $|e\rangle$. In the meantime, the ground states $|0\rangle$ and $|1\rangle$ can also be pumped upwards to the Rydberg state $|r\rangle$ under the actions of two independent classical fields with the same Rabi frequency $\Omega_r$, and commonly detuned by $-\Delta$. For the sake of convenience, we have assumed the excited state $|e\rangle$~($|r\rangle$) spontaneously decays downwards to $|0\rangle$ and $|1\rangle$ with the same rate $\gamma_e/2$~$(\gamma_r/2)$, respectively. Note that the atom-dependent light shift $\delta_i$ plays an important role in the process of $Z$ pumping.}
\end{figure}

We consider three four-level atoms of double $\Lambda$ configuration interact with an optical cavity, and are simultaneously driven by classical laser fields, as shown in Fig.~\ref{fig1}.
 The two stable ground states $|0\rangle$ and $|1\rangle$ are used to be encoded quantum bits. The transition between states $|0\rangle$ and $|e\rangle$ is coupled to a quantized cavity mode with strength $g$, while the transition between states $|1\rangle$ and $|e\rangle$ is coupled to the classical field with Rabi frequency $\Omega$. In the meantime, the ground states $|0\rangle$ and $|1\rangle$ can also be pumped upwards to the excited Rydberg state $|r\rangle$ via two independent classical fields with the same Rabi frequency $\Omega_r$ (generally accomplished by a two-photon process or a direct single-photon process, see Refs.~\cite{PhysRevLett.93.063001,PhysRevA.87.032519,PhysRevA.89.033416} for details), and detuning $-\Delta$. Although it is not necessary, for the sake of convenience we have assumed the excited state $|e\rangle$~($|r\rangle$) can spontaneously decay downwards to $|0\rangle$ and $|1\rangle$ with the same rate $\gamma_e/2$~$(\gamma_r/2)$, respectively. In addition, the atom-dependent light shift $\delta_i$ of state $|1\rangle$ is introduced so as to break the symmetry of ground states during the $Z$ pumping.

 Under the assumption of Markovian approximation, the decay channels for atoms and cavity are independent, thus the master equation describing the interaction between quantum systems and external environment can be modeled by the Lindblad form
\begin{eqnarray}\label{MastEqn1}
\dot{\rho}&=&-i[H_I,\rho]+\frac{\gamma_e}{2}\sum_{j=1}^{3}\big\{{\cal D}[|0\rangle_{jj}\langle e|]\rho+{\cal D}[|1\rangle_{jj}\langle e|]\rho\big\}\nonumber\\&&+\frac{\gamma_r}{2}\sum_{j=1}^{3}\big\{{\cal D}[|0\rangle_{jj}\langle r|]\rho+{\cal D}[|1\rangle_{jj}\langle r|]\rho\big\}+\kappa{\cal D}[a]\rho,
\end{eqnarray}
where $\kappa$ denotes the leaky rate of photon from the optical cavity, ${\cal D}[c]\rho=c\rho c^{\dag}-(c^{\dag}c\rho+\rho c^{\dag}c)/2$ represents the superoperator characterizing decay of system, and the corresponding Hamiltonian $H_I$ reads ($\hbar=1$)
\bea\label{e1}
H_I&=&H_k+H_r,\\
H_k &=& \sum_{i=1}^3\big(\Omega\ket{e}_{ii}\bra{1} +g\ket{e}_{ii}\bra{0}a+{\rm H.c.}+\delta_i|1\rangle_{ii}\langle 1|\big),\nonumber\\
H_r&=&\sum_{i=1}^3\big(\Omega_{r}\ket{r}_{ii}\bra{0}+\Omega_r\ket{r}_{ii}\bra{1}+ {\rm H.c.}-\Delta\ket{r}_{ii}\bra{r}\big)\nonumber\\&&+\sum_{i\neq j}U_{ij}\ket{rr}_{ij}\bra{rr}.\nonumber\eea

It is worth pointing out that there are many ways to implement the atom-dependent light shifts $\delta_i$. For example, these terms can be considered as an extra Stark shift of level $|1\rangle$ via introducing other auxiliary levels (an inverse method adopted generally for canceling the Stark shifts), or an energy difference in a rotating frame through replacing the detuning parameters of the classical field $\Omega$ driving the transition $|1\rangle\leftrightarrow|e\rangle$ and the classical field $\Omega_r$ driving the transition $|1\rangle\leftrightarrow|r\rangle$ by $-\delta_i$ and $-(\Delta+\delta_i)$, respectively.
The Rydberg-mediated interaction $U_{ij}$ originates from the
dipole-dipole potential of the scale $U_{ij}=D(1-3\cos^2{\theta_{ij}})|{\bf R}_i-{\bf R}_j|^{-3}$ between two atoms located at position ${\bf R}_i$ and ${\bf R}_j$, and $\theta_{ij}$ is
the angle between the vector ${\bf R}_i-{\bf R}_j$ and the dipole moment aligned parallel to the $z$ axis, $D=d^2_0/(4\pi\epsilon_0)$, $d_0=(3/2)ea_0n(n-1)$ with $a_0$ the Bohr radius, $e$ the electron charge, and $n$ the principle quantum number \cite{Saffman2010,PhysRevLett.104.223002}.
\section{Simplified $Z$-pumping process}
Let us first investigate the realization of the full $Z$-pumping process by the spontaneous emission of excited state $|e\rangle$ combined with Hamiltonian $H_k$. To make an analogy with the standard quantum Zeno dynamics of Ref.~\cite{Facchi2002,Facchi2008}, we divide the Hamiltonian $H_k$ into two parts, i.e., $H_k=H_0+gH_m$, where $H_0=\Sigma_{i=1}^3(\Omega\ket{e}_{ii}\bra{1}+{\rm H.c.}+\delta_i|1\rangle_{ii}\langle 1|)$ is the interaction between atoms and classical fields, and $gH_m=\Sigma_{i=1}^3g(\ket{e}_{ii}\bra{0}a+{\rm H.c.})$ is the interaction between atoms and cavity. In the limit of $\{|\Omega|,|\delta_i|\}\ll g$, the requirement of quantum Zeno dynamics is fulfilled, and the Hamiltonian is reduced to $H_k=\Sigma_n(P_nH_0P_n+g\eta_nP_n)$ with $P_n$ the orthogonal projection corresponding to the eigenvalue $\eta_n$ of $H_m$. In the Zeno subspace of $\eta_0=0$, it is reasonable to neglect the high-frequency oscillatory terms and only keep the near-resonant transitions, then we have the effective Hamiltonian as follows \cite{Facchi2002,PhysRevA.80.062323,PhysRevLett.117.140502}
\begin{eqnarray}\label{e3}
H^{\rm eff}_k&=&(H^s_k+H^b_k)\otimes|0_c\rangle\langle 0_c|,\
\end{eqnarray}
with
\begin{eqnarray}\label{e4}
H^s_k&=&\Omega\big[|001\rangle\big(\frac{1}{\sqrt{6}}\langle D_1|-\frac{1}{\sqrt{2}}\langle D_2|\big)e^{-i\delta t}\big]\nonumber\\&&+\Omega\big[|100\rangle\big(\frac{1}{\sqrt{6}}\langle D_1|+\frac{1}{\sqrt{2}}\langle D_2|\big)e^{-i\delta t}\big]\nonumber\\&&-\frac{2\Omega}{\sqrt{6}}|010\rangle\langle D_1|e^{2i\delta t}+{\rm H.c.},
\end{eqnarray}
and
\begin{eqnarray}\label{e5}
H^b_k&=&-\frac{\Omega}{\sqrt{2}}\big[|011\rangle\big(\langle D_4|e^{-i\delta t}+\langle D_5|e^{2i\delta t}\big)\big]\nonumber\\&&-\frac{\Omega}{\sqrt{2}}\big[|101\rangle\big(\langle D_3|e^{-i\delta t}-\langle D_5|e^{-i\delta t}\big)\big]\nonumber\\&&+\frac{\Omega}{\sqrt{2}}\big[|110\rangle\big(\langle D_3|e^{2i\delta t}+\langle D_4|e^{-i\delta t}\big)\big]+{\rm H.c.}.
\end{eqnarray}
In the above expressions we have assumed $\delta_1=\delta_3=-\delta=-\delta_2/2$, and this setting will induce additional light shift for each ground state except $|000\rangle$ and $|111\rangle$. The cavity mode is frozen to its vacuum state $|0_c\rangle$ in this subspace, thus the process of $Z$ pumping is robust against the cavity decay. The qubit basis $|ijk\rangle|0_c\rangle~(i,j,k=0,1)$, as well as the quantum states
$|D_1\rangle|0_c\rangle=(|e00\rangle+|00e\rangle-2|0e0\rangle)|0_c\rangle/\sqrt{6}$, $|D_2\rangle|0_c\rangle=(|e00\rangle-|00e\rangle)|0_c\rangle/\sqrt{2}$, $|D_3\rangle|0_c\rangle=(|1e0\rangle-|10e\rangle)|0_c\rangle/\sqrt{2}$, $|D_4\rangle|0_c\rangle=(|e10\rangle-|01e\rangle)|0_c\rangle/\sqrt{2}$, $|D_5\rangle|0_c\rangle=(|e01\rangle-|0e1\rangle)|0_c\rangle/\sqrt{2}$ are the dark states of the atom-cavity interacting Hamiltonian. After discarding the symbol of the cavity field, we obtain the effective Markovian master equation describing the $Z$-pumping process
\begin{equation}\label{MastEqn}
\dot{\rho} = -i[H^{\rm eff}_k,\rho]+\sum_{j=1}^{16} L_j \rho L_j^\dag -\frac{1}{2}(L_j^\dag L_j \rho + \rho L_j^\dag L_j),
\end{equation}
where the Lindblad operator $L_j$ $\in$
$\{\sqrt{\gamma_e/12}|001\rangle\langle D_1|$, $\sqrt{\gamma_e/12}|100\rangle\langle D_1|$, $\sqrt{\gamma_e/3}|010\rangle\langle D_1|$, $\sqrt{\gamma_e/2}|000\rangle\langle D_1|$, $\sqrt{\gamma_e/4}|001\rangle\langle D_2|$, $\sqrt{\gamma_e/4}|100\rangle\langle D_2|$, $\sqrt{\gamma_e/2}|000\rangle\langle D_2|$, $\sqrt{\gamma_e/4}|110\rangle\langle D_3|$, $\sqrt{\gamma_e/4}|101\rangle\langle D_3|$, $\sqrt{\gamma_e/2}|100\rangle\langle D_3|$, $\sqrt{\gamma_e/4}|110\rangle\langle D_4|$, $\sqrt{\gamma_e/4}|011\rangle\langle D_4|$, $\sqrt{\gamma_e/2}|010\rangle\langle D_4|$, $\sqrt{\gamma_e/4}|101\rangle\langle D_5|$, $\sqrt{\gamma_e/4}|011\rangle\langle D_5|$, $\sqrt{\gamma_e/2}|001\rangle\langle D_5|\}$.
\begin{figure}
\scalebox{0.13}{\includegraphics{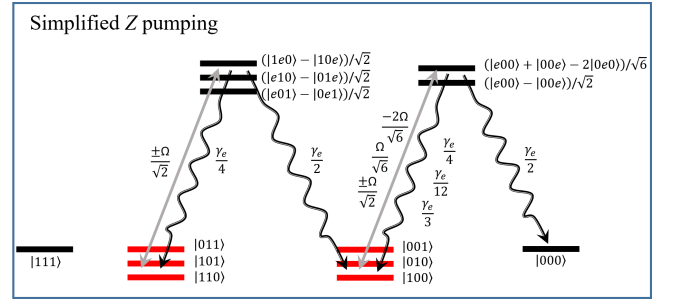} }
\caption{\label{fig2}The effective transitions of quantum states for the simplified $Z$ pumping. The evolution of quantum states is frozen in a Zeno subspace corresponding to the vacuum state of the cavity field. The quantum states with one and two atoms in state $|1\rangle$ are coupled to the excited states by the near-resonant classical fields, and the population of state $|000\rangle$ is accumulated asymptotically to a steady value due to the spontaneous emission of excited states and the coherent pumping of laser fields. During the process, the state $|111\rangle$ is unaffected by the weak driving fields because of the Zeno requirement $\Omega\ll g$.}
\end{figure}

The simplified $Z$-pumping process of our scheme is shown in  Fig.~\ref{fig2}. To be more specific, suppose a quantum state is initialized in $|011\rangle|0_c\rangle$,  it can be first driven into the excited state $|D_4\rangle|0_c\rangle$ with the weak coupling strength $-\Omega/\sqrt{2}$, as governed by the Hamiltonian of Eq.~(\ref{e5}). The excited state $|D_{4}\rangle|0_c\rangle$ then spontaneously decay back to the subspace with two atoms in state $|1\rangle$, i.e. $|011\rangle|0_c\rangle$ and $|110\rangle|0_c\rangle$ with the same emitting rate $\gamma_e/4$ respectively, or to the ground state $|010\rangle|0_c\rangle$ with the emitting rate $\gamma_e/2$, i.e., a quantum state with one atom in state $|1\rangle$. Consider this quantum state $|010\rangle|0_c\rangle$ as a new initial state and repeat a similar pumping and decaying process, the whole system will be finally stabilized into the state $|000\rangle|0_c\rangle$. In general, starting from an arbitrary quantum state with one or two atoms in state $|1\rangle$, the steady state $|000\rangle|0_c\rangle$ is always achievable. As for the ground state $|111\rangle|0_c\rangle$, it is not affected by the above dissipative dynamics because the limit of quantum Zeno dynamics $\Omega\ll g$ contributes an interaction strength at the order of magnitude ${\cal O}(\Omega^2/g)$, which is much smaller than $\Omega$. Now we finish the process of $Z$ pumping with only one classical field acting on atoms and a single-mode cavity. What is more, the cavity mode is not populated throughout the process, making it insensitive to the leakage of photon from the cavity.

\begin{figure}
\scalebox{0.25}{\includegraphics{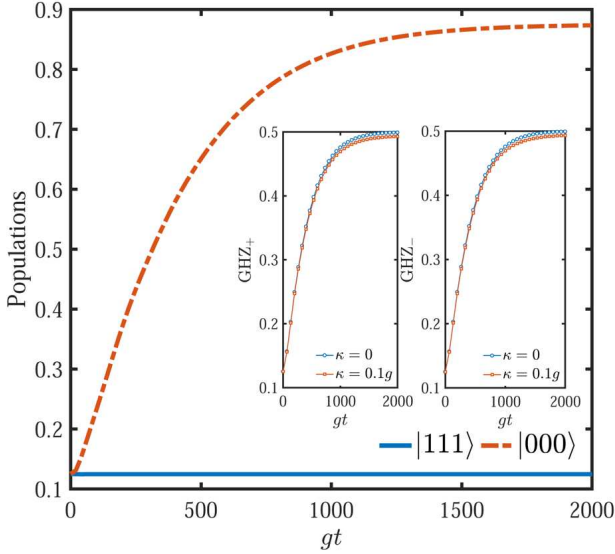} }
\caption{\label{fig3}Numerical simulation of the $Z$-pumping process. The initial state is a fully mixed state in the basis of quantum bits: $\rho_0=\Sigma_{i,j,k=0,1}|ijk\rangle\langle ijk|/8$, and the corresponding parameters are set as $\Omega=0.02g$, $\delta_1=\delta_3=-0.01g$, $\delta_2=0.02g$, $\gamma_e=0.1g$, and $\kappa=0$. The final state is stabilized into $\rho_s=7/8|000\rangle\langle000|+1/8|111\rangle\langle111|$ after a short time $t=2000/g$.}
\end{figure}

In Fig.~\ref{fig3}, we numerically simulate the $Z$-pumping operation with the full Hamiltonian $H_k$ in Eq.~(\ref{e1}). The initial state is chosen as a fully mixed state in the basis of quantum bits: $\rho_0=\Sigma_{i,j,k=0,1}|ijk\rangle\langle ijk|/8$, and the corresponding parameters are set as $\Omega=0.02g$, $\delta_1=\delta_3=-0.01g$, $\delta_2=0.02g$, $\gamma_e=0.1g$, and $\kappa=0$. The population of state $|111\rangle$ (solid line) is invariant and the population of state $|000\rangle$ is stabilized at 0.875 (dash-dotted line) after a relaxation time $t=2000/g$. At this stage, we are  able to prepare a steady GHZ state by a subsequent
quantum feedback operation \cite{PhysRevLett.106.020504,PhysRevLett.115.200502}. A parity check ${\cal P}=\Pi_{i=1}^3(|1\rangle_{ii}\langle0|+|0\rangle_{ii}\langle1|)$ performed on the system can inform us whether the quantum state is ${|\rm GHZ_+}\rangle$ (${\cal P}=1$) or ${|\rm GHZ_-}\rangle$ (${\cal P}=-1$). If the target state is supposed to be ${|\rm GHZ_-}\rangle$ but we acquire a signal of ${\cal P}=1$, a $\sigma_z$ operation applied to one of the qubits will change the state ${|\rm GHZ_+}\rangle$ into the target state ${|\rm GHZ_-}\rangle$. In this sense, the $Z$-pumping operation combined with the parity measurements makes the current proposal deterministic. In the inset of Fig.~\ref{fig3}, we study the evolutions of populations of states ${|\rm GHZ_+}\rangle$ and ${|\rm GHZ_-}\rangle$ in the presence of a large cavity decay ($\kappa=0.1g$). Compared with the ideal case $P=P_{\rm GHZ_+}+P_{\rm GHZ_-}=99.82\%$, although the population of target state is decreased, it remains $98.66\%$. To sum up, we have implemented a robust $Z$-pumping operation.

\section{dissipative Rydberg pumping}
Next we turn to the realization of the dissipative Rydberg pumping. To see this process clearly, we rewrite the Hamiltonian
$H_r=\sum_{i=1}^3(\sqrt{2}\Omega_{r}\ket{r}_{ii}\bra{+}+ {\rm H.c.}-\Delta\ket{r}_{ii}\bra{r})+\sum_{i\neq j}U\ket{rr}_{ij}\bra{rr}$, where we have introduced $|+\rangle=(|0\rangle+|1\rangle)/\sqrt{2}$ and assumed $U_{ij}=U$. Now this model is equivalent to three two-level Rydberg atoms with ground state $|+\rangle$ and excited state $|r\rangle$ collectively driven by a classical field of Rabi frequency $\sqrt{2}\Omega_r$. Using the basis $\{|+++\rangle, (|++r\rangle+|+r+\rangle+|r++\rangle)/\sqrt{3},(|rr+\rangle+|r+r\rangle+|+rr\rangle)/\sqrt{3}, |rrr\rangle\}$, we can reduce the $8\times8$ matrix to a $4\times4$ matrix,
 \begin{eqnarray}\label{one11}
{H}_r=\left[\begin{array}{c c c c}
0 & \sqrt{6}\Omega_{r} &0 & 0 \\
\sqrt{6}\Omega_{r} & -\Delta & 2\sqrt{2}\Omega_{r} & 0  \\
0 & 2\sqrt{2}\Omega_{r} & U-2\Delta &\sqrt{6}\Omega_{r}  \\
0 & 0 &\sqrt{6}\Omega_{r}  & 3U-3\Delta \\
\end{array}
\right].
\end{eqnarray}

In this subspace, a general wave function of quantum system is described by $|\Psi(t)\rangle=c_0(t)|r^{\otimes0}\rangle+c_1(t)|r^{\otimes1}\rangle+c_2(t)|r^{\otimes2}\rangle+c_3(t)|r^{\otimes3}\rangle$, where $|r^{\otimes m}\rangle$ is short for the symmetric state with $m$ atoms in $|r\rangle$. The equations of motion for the probability amplitudes can be derived from the Schr\"{o}dinger equation $i|\dot\Psi\rangle=H_r|\Psi\rangle$ to be
\begin{eqnarray}
i\dot c_0&=&\sqrt{6}\Omega_r c_1,\label{8}\\
i\dot c_1&=&\sqrt{6}\Omega_r c_0+2\sqrt{2}\Omega_rc_2-\Delta c_1,\\
i\dot c_2&=&\sqrt{6}\Omega_r c_1+2\sqrt{2}\Omega_rc_3-\Delta c_2,\\
i\dot c_3&=&\sqrt{6}\Omega_r c_2\label{11},
\end{eqnarray}
and we have set $U=\Delta$. In the limit of $\Delta\gg\Omega_r$, $c_1(t)$ and $c_2(t)$ are slowly varying functions of $t$, thus it is reasonable to assume that $\dot c_1=0$ and $\dot c_2=0$, and acquire the values of these coefficients as
\begin{eqnarray}
c_1&=&\frac{\sqrt{6}\Omega_r}{\Delta} c_0+\frac{2\sqrt{2}\Omega_r}{\Delta}c_2,\\
c_2&=&\frac{\sqrt{6}\Omega_r}{\Delta} c_1+\frac{2\sqrt{2}\Omega_r}{\Delta}c_3.
\end{eqnarray}
By substituting the above results into Eqs.~(\ref{8}) and (\ref{11}), we have a pair of coupled equations characterizing the interaction between states $|+++\rangle$ and $|rrr\rangle$, i.e.,
\begin{eqnarray}
i\dot c_0&=&(\frac{6\Omega_r^2}{\Delta} c_0+\frac{12\sqrt{2}\Omega_r^3}{\Delta^2}c_3)/(1-\frac{8\Omega_r^2}{\Delta^2}),\\
i\dot c_3&=&(\frac{6\Omega_r^2}{\Delta} c_3+\frac{12\sqrt{2}\Omega_r^3}{\Delta^2}c_0)/(1-\frac{8\Omega_r^2}{\Delta^2}),
\end{eqnarray}
which just correspond to the effective Hamiltonian
\begin{equation}\label{rp}
H^{\rm eff}_{r}=\frac{12\sqrt{2}\Omega^3_r}{\Delta^2}|+++\rangle\langle rrr|+{\rm H.c.}.
\end{equation}
The Stark-shift terms have been disregarded in this process since they  can be canceled by introducing ancillary levels, and the order of ${\cal O}(\Omega_r^2/\Delta^2)$ is ignored too. Under the action of the Rydberg pumping of Eq.~(\ref{rp}) and the spontaneous emission of  excited Rydberg states $|r\rangle\xrightarrow{\gamma_r/2}|0(1)\rangle$, the state $|\rm{GHZ_+}\rangle=(|+++\rangle+|+--\rangle+|-+-\rangle+|--+\rangle)/2$ is no longer stable, and it will be pumped and independently decay to the ``bare" ground states. In fact, engineering the coupling between $|rrr\rangle$ and any component of $|\rm{GHZ_+}\rangle$, such as $|+--\rangle$, can achieve the same effect. In other words, pumping the whole state $|\rm{GHZ_+}\rangle$ to the excited state is not necessary \cite{PhysRevLett.117.040501}.

\section{experimental feasibility}
In experiment, we may employ $^{87}$Rb atoms in our proposal. The range of the coupling strength between the atomic transition and the cavity mode is measured from the weak-coupling regime $2\pi\times4.5$~MHz to the strong-coupling regime $2\pi\times215$~MHz \cite{mucke,brennecke,murch,PhysRevLett.100.050401,volz,PhysRevLett.104.203602}. Specifically, a single atom cavity coupling strength is \cite{0953-4075-49-6-064014}
\begin{equation}
g=\mu\sqrt{\frac{\omega_c}{2\epsilon_0V[L_c,R_c\lambda]}},
\end{equation}
where $\mu$ is the atomic transition dipole moment, $V$ is the mode volume of
the cavity, $\omega_c$ is the frequency of cavity, $R_c$ is the radius of curvature of the mirrors, $L_c$ is the cavity
length, and $\lambda$ is the wavelength of the cavity mode. Thus this strength is adjustable by modulating the relevant cavity parameters. The Rabi frequency $\Omega_r$ can be tuned continuously between $2\pi\times(0,100)$~MHz (e.g., a red- and a blue-detuned lasers
on the $5S-5P$ and $5P-|r\rangle$ transitions). The fidelity of the steady state is calculated as $F(\sigma,\rho_{\infty})\equiv{\rm Tr}\sqrt{\sigma^{1/2}\rho_{\infty}\sigma^{1/2}}$, where $\sigma$ is the density matrix of target state. For a pure target state ($\sigma=|s\rangle\langle s|$), the definition of the fidelity can be proved to be $\sqrt{\langle s|\rho_{\infty}|s\rangle}=\sqrt{P}$, which is the square root of the population.

\begin{figure}
\scalebox{0.25}{\includegraphics{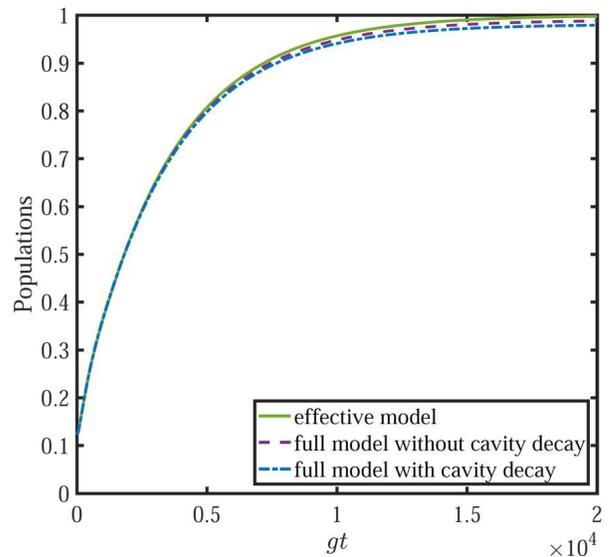} }
\caption{\label{fig4} Numerical simulation of the population for the GHZ state starting from the fully mixed state. The effective master equation (solid line) and the full master equation (dashed line and dash-dotted line) are both simulated with the experimentally achievable parameters: $g=2\pi\times50$~MHz, $\Omega=0.01g$, $\Omega_r=g$, $\delta_1=\delta_3=-0.005g$, $\delta_2=0.01g$, $U=\Delta=58g$, $\kappa=2\pi\times1$~MHz, $\gamma_e=2\pi\times3$~MHz, and $\gamma_r=2\pi\times0.144$~MHz. }
\end{figure}

The experiment of cavity QED with a Bose-Einstein condensate provides us the following parameters
$(g_0,\kappa,\gamma_e)=2\pi\times(10.6,1.3,3)$~MHz \cite{brennecke}. For this group of parameters, we can choose $\Omega=0.002g$, $\delta_1=\delta_3=-0.5\Omega$, $\delta_2=\Omega$, $U=\Delta=100g$, $\Omega_r=g$, and adopt the Rydberg state $|r\rangle=|95D_{5/2};F=4\rangle$ with decay rate $\gamma_r=2\pi\times0.03$~MHz. By substituting these parameters into the original master equation of Eq.~(\ref{MastEqn1}), we obtain the steady-state fidelity $F=96.28\%$, and this value can be further improved to $F=98.15\%$ using another group of experimental parameters
$(g_0,\kappa,\gamma_e)=2\pi\times(14.4,0.66,3)$~MHz \cite{murch}, $\Omega=0.005g$, $\Delta=80g$. In Ref.~\cite{volz}, a fibre-based high-finesse cavity also offers us a set of strong coupling parameters
$(g_0,\kappa,\gamma_e)=2\pi\times(185,53,3)$~MHz. In this condition, the parameter values  $\Omega=0.002g$, $\Delta=40g$, $\Omega_r=0.5g$, and $\gamma_r=2\pi\times0.144$~MHz ($20D$ Rydberg states, see e.g., \cite{PhysRevA.79.052504}) guarantees a high fidelity $F=98.24\%$. To see clearly how fast the system approaches to the steady state $|{\rm GHZ}_-\rangle$ from an arbitrary initial state, we investigate the dependence of the steady-state population on $t$ in Fig.~\ref{fig4}. The solid line, the dashed line and the dash-dotted line are simulated by the effective master equation, the full master equation without and with considering the cavity decay, respectively. These three lines are
in excellent agreement with each other under the given
parameters, which confirms the efficiency of our scheme again. It should be noted that the assumption of identical atom-cavity coupling strength $g$ made throughout the text is only for the discussion convenience. In fact, the fluctuations of $g$ result in little variation in the target-state fidelity. For example, the parameters listed in Fig.~\ref{fig4} corresponds to a steady-state fidelity $99.05\%$. If we replace $g_{1(2,3)}/(2\pi)=50$~MHz with $g_{1}/(2\pi)=50$~MHz, $g_{2}/(2\pi)=45$~MHz, and $g_{3}/(2\pi)=40$~MHz or 55~MHz, the fidelity is still no less than $99.00\%$.

\section{summary}
In summary, we have proposed an efficient mechanism for dissipative generation of the tripartite GHZ state in a Rydberg-atom-cavity QED system. This scheme actively exploits the spontaneous emission of atoms and coherently driving  offered by the quantum Zeno dynamics and the Rydberg pumping, which make it robust against the loss of cavity and the fluctuation of atom-cavity couplings. Although the current model is not scalable, it enables us to reduces the operation complexity of the experiment substantially, and a high fidelity is available through the strictly numerical simulation of the full master equation without any approximation.  We hope that our proposal may open a new venue for the experimental
realization of the multipartite entanglement in the near future.

\section{Acknowledgements}
This work is supported by the Natural Science Foundation
of China under Grants No. 11647308, No. 11674049,
No. 11534002, and No. 61475033, No. 11774047, and by Fundamental
Research Funds for the Central Universities under Grant No.
2412016KJ004.

\appendix*
\section{DETAILED DERIVATION OF THE ZENO HAMILTONIAN FOR THE $Z$ PUMPING}
In this appendix, we give the detailed derivation of the effective Hamiltonian of Eq.~(\ref{e4}).
For the qubit states with one atom in state $|1\rangle$, we can obtain a closed subspace $\{|001\rangle|0_c\rangle$, $|010\rangle|0_c\rangle$, $|100\rangle|0_c\rangle$,
$|00e\rangle|0_c\rangle$, $|0e0\rangle|0_c\rangle$, $|e00\rangle|0_c\rangle$,
$|000\rangle|1_c\rangle\}$ in the absence of dissipation. Now we expand the original Hamiltonian $H_k$ in Eq.~(\ref{e1}) with the above basis and have
\begin{eqnarray}\label{e44}
H_0^{ap}&=&\Omega[|001\rangle\langle00e|+|010\rangle\langle0e0|+|100\rangle\langle e00|+{\rm H.c.}\nonumber\\&&-\delta(|1\rangle_{11}\langle1|-2|1\rangle_{22}\langle1|+|1\rangle_{33}\langle1|)]|0_c\rangle\langle0_c|,
\end{eqnarray}
and
\begin{equation}\label{e45}
H_g^{ap}=g[(|00e\rangle+|0e0\rangle+|e00\rangle)\langle000|]|0_c\rangle\langle1_c|+{\rm H.c.},
\end{equation}
where $H_0^{ap}$ and $H_g^{ap}$ represent the interactions between atoms and classical fields, and atoms and cavity, respectively. According to the Zeno dynamics \cite{Facchi2002,Facchi2008}, we should first find the eigenprojections of $H_g^{ap}$. After a straightforward calculation, we get four eigenstates of $H_g^{ap}$ as
\begin{equation}\label{e46}
|E_1\rangle=\frac{1}{\sqrt{6}}(|e00\rangle+|00e\rangle-2|0e0\rangle)|0_c\rangle,
\end{equation}
\begin{equation}\label{e46}
|E_2\rangle=\frac{1}{\sqrt{2}}(|e00\rangle-|00e\rangle)|0_c\rangle,
\end{equation}
\begin{equation}\label{e46}
|E_3\rangle=\frac{1}{\sqrt{6}}(|e00\rangle+|0e0\rangle+|00e\rangle)|0_c\rangle+\frac{1}{\sqrt{2}}|000\rangle|1_c\rangle,
\end{equation}
\begin{equation}\label{e46}
|E_4\rangle=\frac{1}{\sqrt{6}}(|e00\rangle+|0e0\rangle+|00e\rangle)|0_c\rangle-\frac{1}{\sqrt{2}}|000\rangle|1_c\rangle,
\end{equation}
corresponding to eigenvalues $0$, $0$, $\sqrt{3}g$, and $-\sqrt{3}g$, respectively. Remember that the qubit states $|\alpha\rangle\in$ $\{|001\rangle|0_c\rangle$, $|010\rangle|0_c\rangle$, $|100\rangle|0_c\rangle\}$ are also the dark states for $H_g^{ap}$ because of $H_g^{ap}|\alpha\rangle=0$, therefore there are total three Zeno subspaces, i.e.,
\begin{eqnarray}\label{e47}
{\cal H}_{p_0}&=&{\rm span}\{|001\rangle|0_c\rangle,|010\rangle|0_c\rangle,|100\rangle|0_c\rangle, |E_1\rangle, |E_2\rangle \},\nonumber\\
{\cal H}_{p_1}&=&{\rm span}\{|E_3\rangle \},\  \  \  \  \  \  \  \  \  \
{\cal H}_{p_2}={\rm span}\{|E_4\rangle \}.
\end{eqnarray}
\begin{figure}
\scalebox{0.5}{\includegraphics{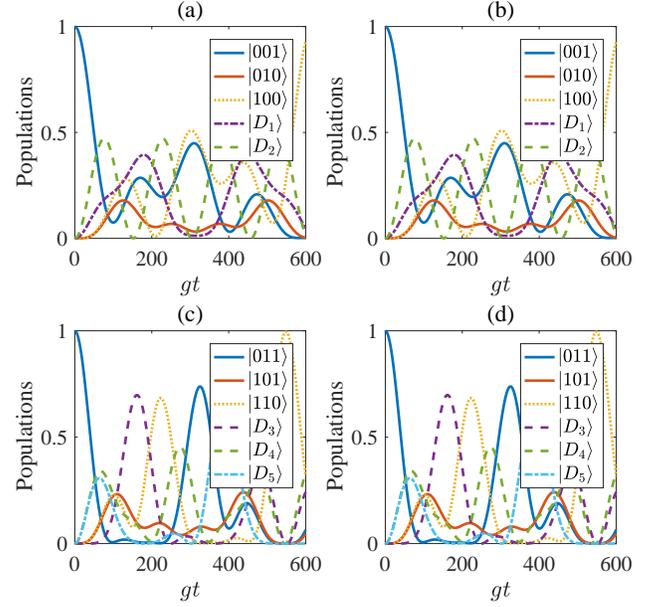} }
\caption{\label{afig4} Numerical simulation of the populations for quantum states using the full Hamiltonian $H_k$ in Eq.~(\ref{e1}) in (a) and (c), compared with the results obtained from utilizing the effective Hamiltonian of Eq.~(\ref{e3}) in (b) and (d). The initial states are chosen as $|001\rangle|0_c\rangle$ and $|011\rangle|0_c\rangle$, respectively, and the corresponding parameters are $\Omega=0.02g$, $\delta_1=\delta_3=-0.01g$, and $\delta_2=0.02g$.}
\end{figure}
Now we rewrite  $H_k$ in the eigenbasis of $H_g^{ap}$ as
\begin{eqnarray}\label{e48}
H_k^{ap}&=&\sum_{m,n=0}^2(P_mH_0P_n+g\eta_nP_n)\nonumber\\
&=&\Omega\big\{|001\rangle|0_c\rangle\big[\frac{1}{\sqrt{6}}(\langle E_1|+\langle E_3|+\langle E_4|)\nonumber\\&&-\frac{1}{\sqrt{2}}\langle E_2|\big]\big\}+\Omega\big\{|100\rangle|0_c\rangle\big[\frac{1}{\sqrt{6}}(\langle E_1|+\langle E_3|\nonumber\\&&+\langle E_4|)+\frac{1}{\sqrt{2}}\langle E_2|\big]\big\}-\Omega|010\rangle|0_c\rangle[\frac{1}{\sqrt{6}}(2\langle E_1|\nonumber\\&&-
\langle E_3|-\langle E_4|)\big]+{\rm H.c.}-\delta(|100\rangle\langle100|\nonumber\\
&&-2|010\rangle\langle010|+|001\rangle\langle001|)|0_c\rangle\langle0_c|
\nonumber\\&&+\sqrt{3}g|E_3\rangle\langle E_3|-\sqrt{3}g|E_4\rangle\langle E_4|.
\end{eqnarray}
In order to see the Zeno dynamics clearly, we move into a rotating frame with respect to $\exp\{-it[\delta(|100\rangle\langle100|-2|010\rangle\langle010|+|001\rangle\langle001|)|0_c\rangle\langle0_c|
+\sqrt{3}g|E_3\rangle\langle E_3|-\sqrt{3}g|E_4\rangle\langle E_4|]\}$ and obtain
\begin{eqnarray}\label{ef}
H_k^{ap}&=&\Omega\big\{|001\rangle|0_c\rangle\big[\big(\frac{1}{\sqrt{6}}\langle E_1|-\frac{1}{\sqrt{2}}\langle E_2|\big)e^{-i\delta t}\nonumber\\&&+\frac{1}{\sqrt{6}}(\langle E_3|e^{-i(\sqrt{3}g+\delta)t}+\langle E_4|e^{i(\sqrt{3}g-\delta)t})\big]\big\}\nonumber\\&&+\Omega\big\{|100\rangle|0_c\rangle\big[\big(\frac{1}{\sqrt{6}}\langle E_1|+\frac{1}{\sqrt{2}}\langle E_2|\big)e^{-i\delta t}\nonumber\\
&&+\frac{1}{\sqrt{6}}(\langle E_3|e^{-i(\sqrt{3}g+\delta)t}+\langle E_4|e^{i(\sqrt{3}g-\delta)t})\big]\big\}\nonumber\\&&-\Omega|010\rangle|0_c\rangle[\frac{1}{\sqrt{6}}(2\langle E_1|e^{2i\delta t}-
\langle E_3|e^{-i(\sqrt{3}g-2\delta)t}\nonumber\\&&-\langle E_4|e^{i(\sqrt{3}g+2\delta)t})\big]+{\rm H.c.}.
\end{eqnarray}
In the limit of Zeno requirement $\{|\Omega|,|\delta|\}\ll g$, the high-frequency
oscillating terms proportional to $\exp[\pm i\sqrt{3}gt]$ can be safely neglected and only the near-resonant terms are preserved. Then we can recover the effective Hamiltonian of Eq.~(\ref{e4}) from Eq.~(\ref{ef}). The effective Hamiltonian of Eq.~(\ref{e5}) can be derived in the same way, where the two-excitation states with two atoms in state $|e\rangle$ are disregarded since the Rabi frequency of the classical fields is weak. In Fig.~\ref{afig4}, we check the effectiveness of Eq.~(\ref{e3}) by plotting the populations for quantum states in Fig.~\ref{afig4}(b) and Fig.~\ref{afig4}(d), and comparing the corresponding results obtained from the full Hamiltonian $H_k$ of Eq.~(\ref{e1}) in Fig.~\ref{afig4}(a) and Fig.~\ref{afig4}(c), which shows that they are in excellent agreement
with each other under the given parameters.

\bibliographystyle{apsrev4-1}
\bibliography{dissipation_GHZ}

\end{document}